\documentclass[twocolumn,amsmath,amssymb, aps,longbibliography]{revtex4-2}
\usepackage{graphicx}
\usepackage{arydshln}
\usepackage{enumitem}
\usepackage{tabularx}
\usepackage[table,xcdraw]{xcolor}
\usepackage{soul} 
\usepackage{graphicx}
\usepackage{dcolumn}
\usepackage{bm}

\begin{document}

\preprint{APS/123-QED}

\title{Structure and Phase Transitions of Metastable Hexagonal Uranium Thin Films}

\author{Rebecca Nicholls}
\email{beckie.nicholls@bristol.ac.uk}
\affiliation{H.H. Wills Physics Laboratory \\ University of Bristol BS8 1TL, UK.}
\author{Chris Bell}
\affiliation{H.H. Wills Physics Laboratory \\ University of Bristol BS8 1TL, UK.}
\author{Johann Bouchet}
\affiliation{CEA, DES, IRESNE, DEC, SESC, LM2C, F-13108 Saint-Paul-Lez-Durance, France}%
\author{Ross Springell}
\affiliation{H.H. Wills Physics Laboratory \\ University of Bristol BS8 1TL, UK.}
\author{Gerard H. Lander}
 \affiliation{H.H. Wills Physics Laboratory \\ University of Bristol BS8 1TL, UK.}

\date{\today}
\begin{abstract}
We report a simple technique for the synthesis of uniaxially textured, metastable hexagonal close-packed-like uranium thin films with thicknesses between $175-2800$ {\AA}. The initial structure and texture of the layers have been studied via X-ray diffraction and reflectivity and the time-dependent transitions of the samples into various orientations of orthorhombic $\alpha$-U have been mapped by similar techniques. The final crystallographic orientations of each system and the timescales on which the transitions occur are found to depend on the lattice parameters of the original layer. The absence of the $\alpha$-U (001) orientation in the transition products suggests that the transitions in these layers are mediated by mechanisms other than the [110] transverse acoustic phonon mode previously suggested for the cubic $\gamma$-U(110) to hcp-U(00.1) to $\alpha$-U(001) displacive phase transition. Alternative transition pathways are discussed.

\end{abstract}

\maketitle


\section{\label{sec:level1}Introduction}
Under ambient conditions, most elements crystallise in high-symmetry structures where an isotropic local environment is provided by the neighbouring atoms e.g. face-centred cubic (fcc), hexagonal close-packed (hcp) and body-centred cubic (bcc) arrangements. In uranium metal, however, the narrow 5$f$ electron bandwidth (1–3 eV) near the Fermi level drives the formation of several uncommon low-symmetry structures in the bulk material. The ground state, orthorhombic $\alpha$-U (Cmcm) structure is not adopted by any other element under ambient pressure and hosts a series of charge density wave states below 43 K, as well as superconductivity below 1 K \cite{Lander1994TheReview}. Upon heating, uranium transforms into the complex tetragonal $\beta$-U structure (P$4_2$/$mnm$) at 935 K, crystallises in the body-centred cubic $\gamma$-U structure (Im3m) at 1045 K and melts at 1405 K. Several computational studies have calculated the energetics and relative stabilities of other theoretical allotropes \cite{Soderlind1998TheoryActinides, Beeler2013FirstUranium, Schonecker2012FerromagneticPrediction} but the most symmetric structures (e.g. hcp and fcc) have never been observed experimentally in bulk, even under pressures of up to 100 GPa \cite{Yoo1998PhaseTemperatures, LeBihan2003StructuralGPa, Dewaele2013Refinement-uranium,Bouchet2017High-temperatureUranium}.   

In contrast, recent work using epitaxial engineering of thin films has shown that strain effects can be exploited to stabilise energetically unfavourable orientations and phases of metallic uranium. For example, the CDW transition temperature in epitaxial $\alpha$-U has been increased to 120 K by straining the $a$-axis \cite{Springell2014MalleabilityFilms}, thin films of pure $\beta$-U have remained stable under ambient conditions for over a year despite its instability in bulk at room temperature \cite{Yang2021MicrostructureFilms} and single crystals of Mo-doped $\gamma$-U, which are yet to be synthesised in bulk, can be epitaxially stabilised on Nb(110) \cite{Chaney2021}. There is also evidence for a fourth `hidden' allotrope of uranium metal that exists only as a thin film. 

The hexagonal phase of U metal was first alluded to when 70 {\AA} of uranium deposited onto single crystal W(110) substrates produced sharp, six-fold symmetric LEED patterns \cite{Berbil-Bautista2004ObservationSpectroscopy,Boysen1998DispersionCeRh3,Molodtsov1998DispersionMetal}. Comparisons of the experimentally and theoretically derived band structures suggested that the pattern is indeed representative of a hexagonal close-packed (P6$_3$/$mmc$) structure, rather than a twinned fcc state \cite{Molodtsov1998DispersionMetal,Chen2019DirectW110, Molodtsov2001InterpretationSystems, Berbil-Bautista2004ObservationSpectroscopy}. However, no ex-situ structural characterisation of these films was performed and the sub-surface structure was not fully determined. Other studies later showed that the lattice parameters of the phase can depend heavily on the both the layer thickness and the underlying crystallographic character of the surface. The lattice parameter ($c/a$) ratios in the few published reports of hcp-U vary from near-ideal close-packed ratios ($1.63$) for $<100$ {\AA} uranium layers grown onto W(110) substrates, up to $c/a = 1.90$ for a 500 {\AA} layer deposited onto a Gd(00.1) buffer layer \cite{Springell2008ElementalFilms}. Recent observations of hexagonal surface reconstructions in epitaxial (001) and (110)-oriented $\alpha$-U thin films complicates the situation further \cite{Nicholls2022ExploringUnpublished}.

The hcp structure observed in the Gd/U system is of interest as the large $c/a$ ratio makes the allotrope one of the most anisotropic elemental hcp structures. Several calculations have predicted that an anisotropic bulk hcp structure would order ferromagnetically and exhibit large spin-Hall angles \cite{Springell2008ElementalFilms, Zarshenas2012TheoreticalProperties,Wu2020Spin-dependentUranium} whereas other studies refute claims of magnetic ordering in bulk and suggest that strained thin films will not be magnetic \cite{Schonecker2012FerromagneticPrediction}. These studies did not take into consideration that the theoretical phonon dispersion for the hcp phase, included in Section II of this paper, predicts that this high symmetry allotrope is dynamically unstable at the $c/a$ value which minimises its total energy.

Experiments which could elucidate the stabilisation mechanics of hcp-U, such as inelastic X-ray scattering, require thicker single crystal layers than those currently achieved in literature. However, the unpredictability of uranium crystallisation and stability on both Gd(00.1) and W(110) is problematic, as highlighted by the growth of unusual multi-domain $\alpha$-U in Ref.~\cite{Ward2008TheUranium} and by the preference for $\alpha$-U in thick (300-3000 {\AA}) layers deposited onto W(110) at growth temperatures of 300-800$^\circ$C \cite{Springell2014MalleabilityFilms,Nicholls2022ExploringUnpublished}. In light of these limitations, it is important that other possible templates for hcp-U growth are investigated to give insight into the wide range of $c/a$ ratios and crystallographic character observed in the literature. 

During the search for alternative buffer layers which promote reliable single crystal growth, we developed a simple technique for the growth of metastable hcp films of uranium metal onto textured Cu$\{111\}$ and Ir$\{111\}$ across a range of thicknesses. Intriguingly, we are able to measure the real-time relaxation of the hcp-like structure into the orthorhombic $\alpha$-U phase. We find that the hcp unit cell parameters, the rate of transformations, and the crystallographic texture of the final $\alpha$-U films are all dependent on the thickness of the uranium layer ($t_U$). These data shed light on the relationship between the hexagonal and orthorhombic phases of elemental uranium, suggesting that the soft phonon mode proposed by Axe is not the driving mechanism for the transition in highly anisotropic hcp structures \cite{Axe1994StructureMetal}. 

The structure of this paper is as follows: Section II shows the theoretical phonon dispersion and lattice parameters of relaxed hcp-U. Section III presents details of the experimental sample growth and basic characterisation. Section IV discusses more detailed X-ray diffraction (XRD) and X-ray reflectivity (XRR) measurements which have been used to study the evolution of the lattice parameters and microstructure with thickness. These techniques are also used in Section V to track the transition of the samples from their original state into $\alpha$-U. Finally, in Section VI we discuss alternative mechanisms for the hcp-U to $\alpha$-U transitions based on the experimental observations.

\section{Instability of hcp-U}
\begin{figure}[t]
\includegraphics[width=0.45\textwidth]{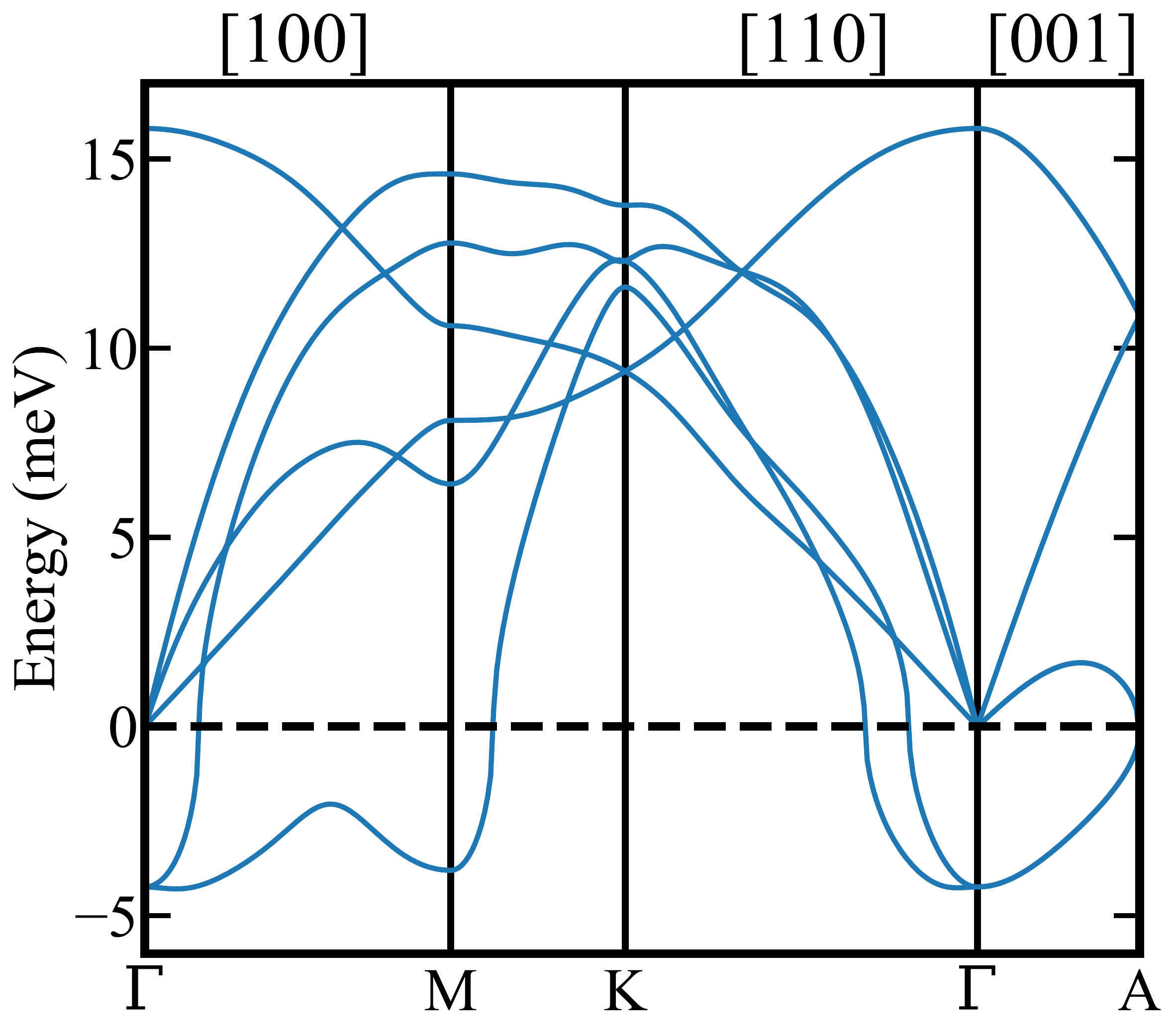}
\caption{\label{fig:PhononDispersion}Calculated phonon dispersion for hcp-U ($c/a=1.84$ and $V=21.2$ {\AA}$^3$). Imaginary phonon energies (plotted as negative) are found near the $\Gamma$ and M points. The volume of bulk $\alpha$-U at room temperature is 20.75 {\AA}$^3$  \cite{Barrett1963CrystalTemperatures}.}
\end{figure}

One of the main motivations for the present effort to stabilise hcp-U was the calculation of the phonon dispersion spectra in Fig.~\ref{fig:PhononDispersion}. \textit{Ab initio} simulations were performed using the \textsc{ABINIT} package \cite{Gonze2020,romero_abinit_2020} in the framework of the Projector Augmented Wave (PAW) method \cite{PAW1,PAW2} and by means of the generalized gradient approximation (GGA) according to the parameterization of Perdew, Burke and Ernzerhof (PBE) for the exchange-correlation energy and potential \cite{Perdew}. Results are obtained using a plane-wave cutoff energy equal to 600 eV. We considered 14 electrons in the valence and a cutoff radius of 2.85 a.u. for U and a $10\times10\times5$ k-points mesh for the hcp structure. Dynamical matrices were calculated for q-points on a similar grid. The relaxation of the hcp structure gives a volume of 21.2 {\AA}$^3$ and a $c/a$ ratio of 1.84, in agreement with calculations performed with the full-potential linear muffin-tin-orbital (FPLMTO) method \cite{Springell2008ElementalFilms}. As expected, the energy of this high symmetry structure exceeds the total energy of $\alpha$-U. A number of imaginary modes in the optical branches (plotted with negative energy in Fig.~\ref{fig:PhononDispersion}) also indicate that hcp-U is dynamically unstable.

\section{Growth and Characterisation}
Thin film samples were grown using a dedicated actinide D.C. magnetron sputtering system (University of Bristol), which is a four-gun assembly housed in an ultra-high vacuum chamber with a base pressure better than $1\times$10$^{-9}$ mbar. All layers were deposited without intentional substrate heating using $7.4-7.6\times$10$^{-3}$ mbar argon gas as the sputtering medium. 

\begin{figure}[h]
\includegraphics[width=0.45\textwidth]{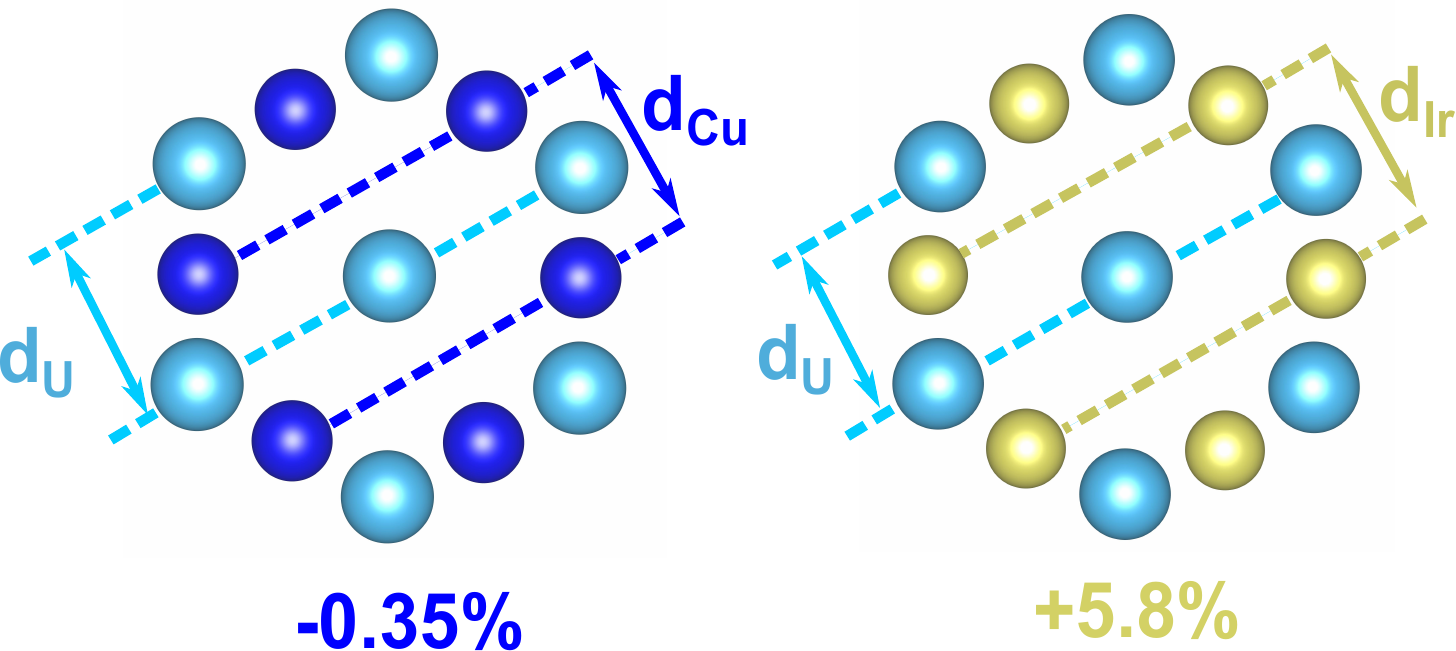}
\caption{\label{fig:match}Schematics of the expected epitaxial matches between hcp-U(00.1) and Cu(111)/Ir(111) using lattice parameters of $a=2.96$ {\AA} and $c=5.625$ {\AA} from Ref.~\cite{Springell2008ElementalFilms}. Values of predicted percentage strains for the U layer are shown (compressive for Cu, tensile for Ir). The real space alignment U[11.0]$\parallel$Cu[11-2] produces a match governed by the distance $d_{\mathrm{U}}=2.563$ {\AA} with $d_{\mathrm{Cu}}=2.565$ {\AA} or $d_{\mathrm{Ir}}=2.715$ {\AA}.}
\end{figure}

Buffer layers of copper or iridium with nominal thicknesses of $t_{\mathrm{Cu}}=175$ {\AA} and $t_{\mathrm{Ir}}=145$ {\AA} were deposited directly onto epi-polished Corning glass substrates (MTI Corporation). Since Cu and Ir are fcc metals, both tend to adopt the close-packed $\{111\}$ orientation when deposited onto an amorphous surface at room temperature, providing a rotationally disordered but locally `hexagonal' template for the uranium overlayer. If the hcp-U film perfectly and coherently strains to the buffer, the Cu template generates $a_{hcp}=2.951$ {\AA} and the Ir template generates $a_{hcp}=3.135$ {\AA}. These parameters `book-end' the range of reliable lattice parameters (2.96(2)-3.2(1) {\AA}) found in literature, and the 500 {\AA} thick hcp-U single crystal from Ref.~\cite{Springell2008ElementalFilms} gives the predicted epitaxial match and strains illustrated in Fig.~\ref{fig:match}.

\begin{figure}[h]
\includegraphics[width=0.45\textwidth]{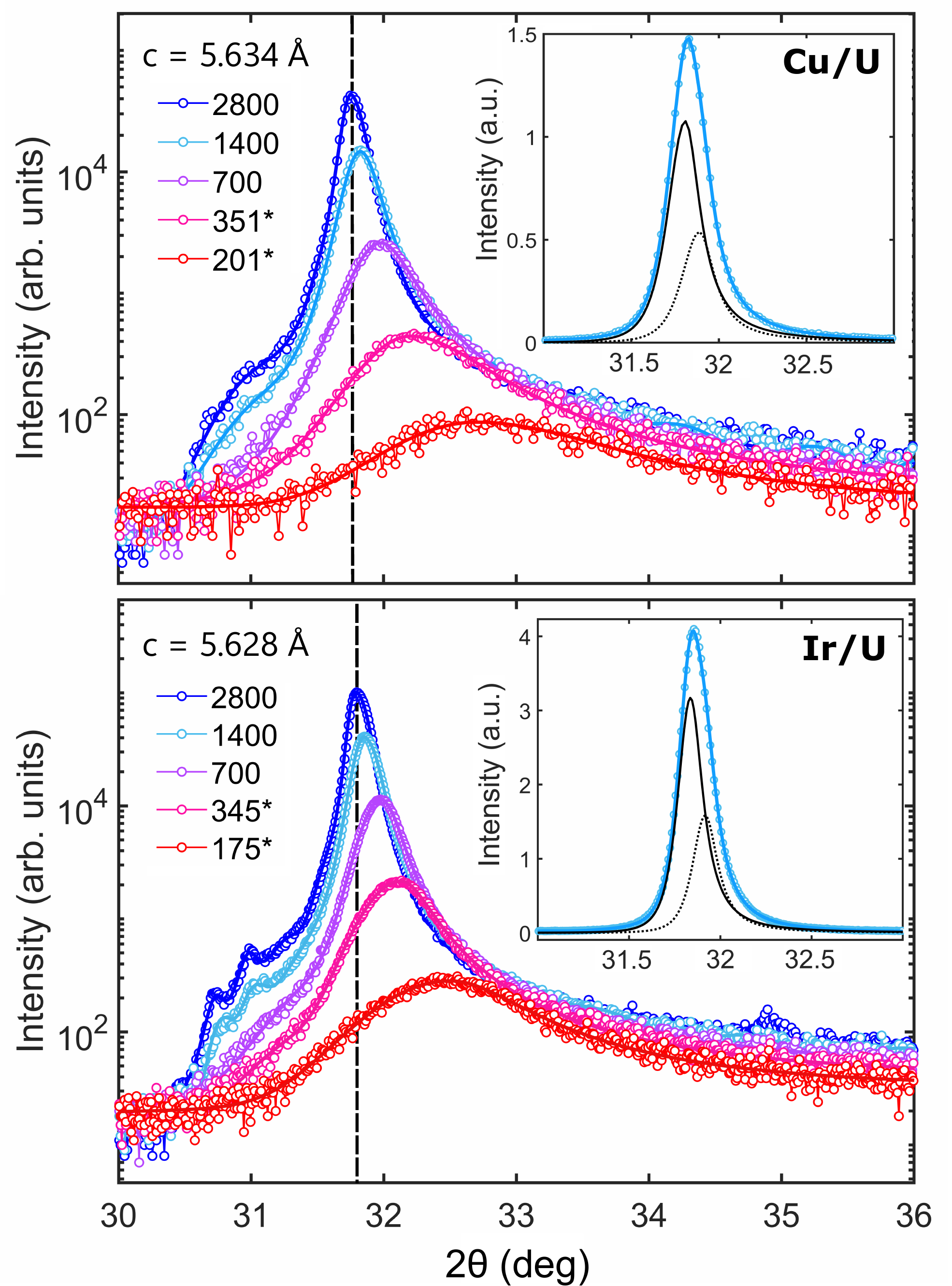}
\caption{\label{fig:002refs}Evolution of the hcp-U (00.2) Bragg reflections for various uranium thicknesses ($t_{\mathrm{U}}$) in Cu/U (top) and Ir/U (bottom) bilayers, labelled with the $c$ parameter of the thickest film (dashed vertical line). Insets show the fit to the 1400 {\AA} reflection on a linear scale, solid and dashed lines correspond to the  Cu-K$\alpha_1$ and Cu-K$\alpha_2$ wavelengths, respectively. Thicknesses labelled (*) were directly derived from X-ray reflectivity profiles. Shoulder regions are attributed to secondary X-rays from the Ni filter in the diffractometer.}
\end{figure}

Uranium metal was deposited onto the $\{111\}$-textured surfaces at room temperature at a rate of 1.4 {\AA}/s to produce layers with nominal thicknesses ($t_{\mathrm{U}}$) of 175, 350, 700, 1400 and 2800 {\AA}, as calibrated by X-ray reflectivity in relatively thin samples. Each bilayer was capped with a layer of Nb (100 {\AA} in the Cu systems, 200 {\AA} in the Ir systems) to prevent oxidation before being rapidly transferred to a Philips X'Pert diffractometer with a Cu-K$_\alpha$ source. The out-of-plane texture of each bilayer was measured through specular (2$\theta$-$\omega)$ and rocking curve ($\omega$) scans. Scans taken within the first hour after growth showed only a series of strong asymmetric reflections characteristic of the $(00.l)$ family ($l=2n$, where $n=1,2,3..$) in an hcp-structure (see Supplemental Material). No other secondary phases or orientations were observed  Similar scans were repeated in the weeks following growth to evaluate the stability of each system.

\begin{figure*}[t]
\includegraphics[width=0.8\textwidth]{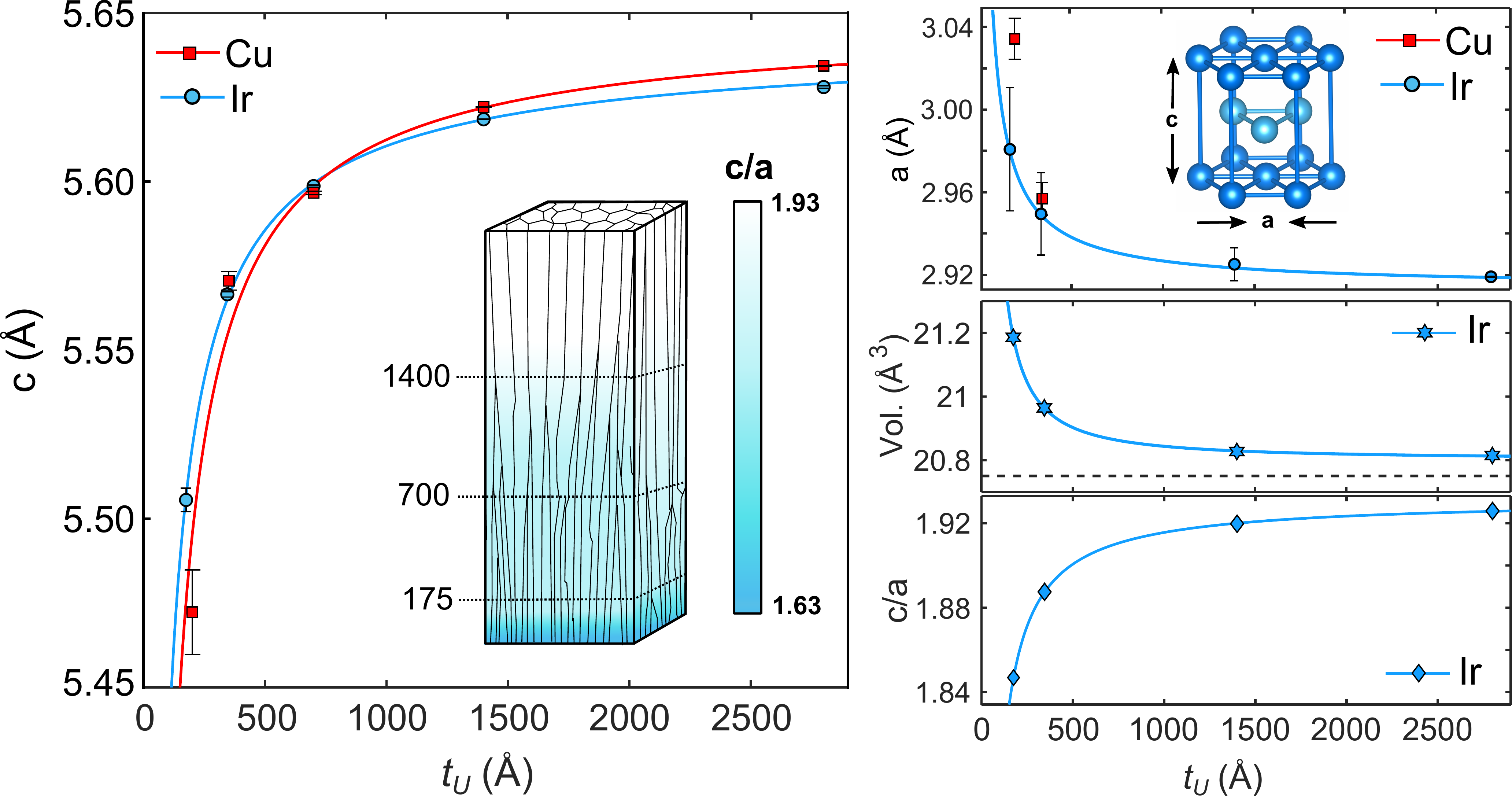}
\caption{\label{fig:IrCu_CValues}Evolution of the hcp-U lattice parameters and atomic volume and in Cu/U and Ir/U bilayers as a function of $t_{\mathrm{U}}$. Left-hand inset is an exaggerated illustration of the simultaneous structural relaxation and columnar merging. Right-hand inset shows an hcp structure with the direction of change in $a$ and $c$ labelled. The dashed line indicates the atomic volume of room temperature, bulk $\alpha$-U (20.75 {\AA}$^3$) from Ref.~\cite{Barrett1963CrystalTemperatures}.}
\end{figure*}

Separate studies were conducted using single crystal Ir(111) and Cu(111) buffer layers grown at relatively high temperatures onto single crystal Al$_2$O$_3$(00.1) and MgO(111), respectively. Subsequent uranium deposition at temperature intervals between RT$-500^\circ$C showed no evidence of single crystal hcp-U formation on either buffer layer. At (nominal) room temperature, an hcp-U layer formed with \{00.1\}-texture but did not adopt any in-plane alignment on either buffer. Higher growth temperatures destroyed the hexagonal state on both the Cu and Ir buffers and, in the case of Ir/U, caused intermixing of the U and Ir layers.

The in-plane lattice parameters in the thickest Corning/Ir/U bilayers were probed by applying offsets to the diffractometer to align an expected hcp off-specular reflection (i$.$e$.$ a Bragg peak with some in-plane component) with the point detector. In a preferred orientation (00.1) system with perfectly random in-plane ordering, the intensity of each off-specular reflection is invariant to rotation about the [00.1] axis. Coupled $2\theta$-$\omega$ scans which intersect this isotropic ring of intensity give an angular (2$\theta$) position from which the in-plane lattice parameters of the layer can be refined. Here we assume an ideal in-plane hexagonal structure, i$.$e$.$ the two sides of the primitive unit cell are the same length ($a=b$) and the angle between them $\gamma=120^\circ$. Though a full structural refinement is beyond the scope of this paper, we were able to find several off-specular peaks in the thicker Ir/U samples that are strongly suggestive of a Mg-type hcp structure. For the $c$-axis oriented 2800 {\AA} Ir/U bilayer, least-squares refinement of the $a$ parameter has been performed using the (10.4), (11.4), (10.5) and (20.4) peak positions (see Supplemental Material for peak positions). For the 1400 {\AA} Ir/U system, the (10.4) and (11.4) positions have been used. The existence of the hcp (11.4) reflection and its position in reciprocal space distinguishes the structure from the textured $\beta$-U(001) films in Ref.~\cite{Yang2021MicrostructureFilms}. The off-specular reflections are relatively weak and were not measured for the Cu/U series.

X-ray reflectivity was also used to determine the sample thickness, electronic density and interfacial roughness of layers in samples with total thickness $<1000$ {\AA}. The reflectivity profiles (see Supplemental Material for profiles and fitting parameters) were fitted using GenX, where error bars on each fitting parameter are calculated from a 5\% change in the optimal figure of merit \cite{Bjorck2007GenX:Evolution}. The $a$ lattice parameters of the thinnest uranium layers were indirectly calculated by combining the XRR-determined electronic densities with the $c$-parameters from the specular reflections, again assuming an ideal hcp structure. The $a$ parameter of the 700 {\AA} Ir-U sample was not determined due to the low intensity off-specular reflections and an overall thickness too close to the XRR limit. 

\section{Hexagonal Thin Films}

As expected from the $\{111\}$-texture of the buffer, the as-deposited uranium forms a textured hcp-like layer with the $c$-axis oriented out of plane. Fig.~\ref{fig:002refs} shows the U(00.2) reflection from each Cu/U and Ir/U bilayer, fitted with asymmetric lineshapes. A plot showing the evolution of the full width at half maximum (FWHM) of the specular peak ($\Delta2\theta$) and the associated rocking curve ($\Delta\omega$) with $t_U$ is included as Supplemental Material. As expected from the Scherrer equation, the width of the specular diffraction peak is inversely proportional to the layer thickness, $t_U$. The sharp decrease and plateauing of the rocking curve width with $t_U$ is suggestive of a microstructural evolution with increasing thickness, as is observed in other metallic thin films systems grown with low adatom mobility. In the context of the zone-dependent models of References~\cite{Messier1984RevisedStructure} and \cite{Thornton1986TheCoatings}, these bilayers were grown in `Zone 1', where $T_{\mathrm{growth}}/T_{\mathrm{melting}}<0.3$ (in Kelvin). Films grown in this regime are characterised by a low density `void' network surrounding an array of narrow, parallel columns which coalesce on a non-linear, thickness-dependent scale. This increasing distance between grain boundaries would generate the observed reduction in $\Delta\omega$. Evidence for coherent column formation in the thin limit is discussed in Section V. We also note that fibrous, columnar structures have been found in textured films of $\beta$-U deposited onto Si(111) at room temperature \cite{Yang2021MicrostructureFilms}.  

Fig.~\ref{fig:IrCu_CValues} shows the evolution of the uranium lattice parameters with increasing layer thickness, where the $1/t_{\mathrm{U}}$ functional forms in both $a$ and $c$ are often seen in thin films undergoing plastic relaxation above a critical thickness \cite{Wang2013CriticalFilms,DunstanGeometricalGrowth}. The 1\% overall reduction in the atomic volume is driven by a simultaneous 2\% contraction in $a$ and 2.2\% expansion in $c$, causing a contraction from $V=21.2$~{\AA}$^3$ to 20.8~{\AA}$^3$. This brings the hcp-U atomic volume closer to bulk $\alpha$-U at 20.75~{\AA}$^3$ \cite{Lander1970NeutronTemperatures} as the $c/a$ ratio increases from 1.85 to 1.93. 

\begin{figure}[t]
\includegraphics[width=0.45\textwidth]{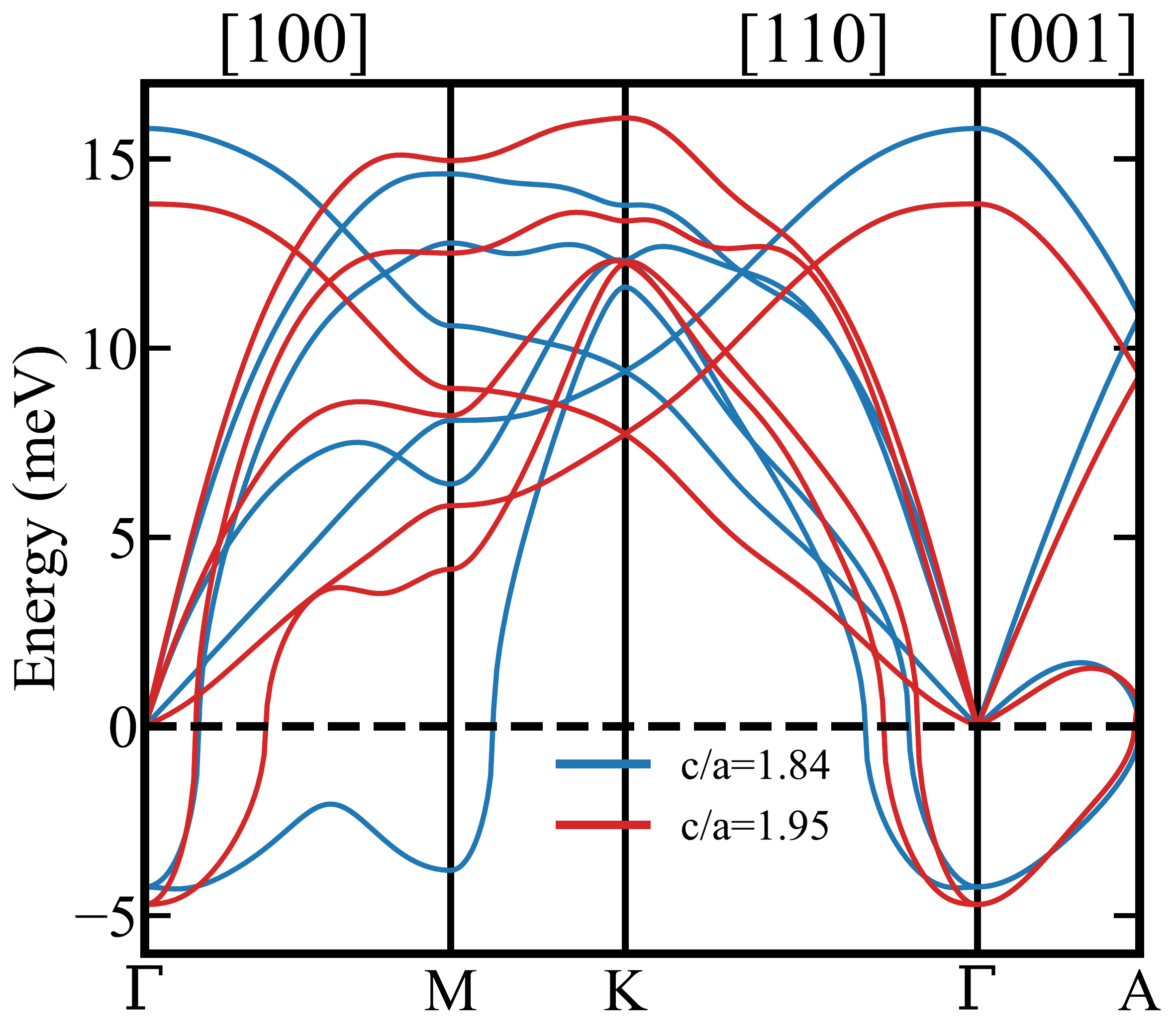}
\caption{\label{fig:PhononDispersionca}Calculated phonon dispersion for hcp-U for $c/a$=1.84 (in blue) and 1.95 (in red) and a volume of $V=21.2$ {\AA}$^3$.}
\end{figure}

Fig.~\ref{fig:PhononDispersionca} then shows the influence of the $c/a$ ratio on the theoretical phonon spectrum of hcp-U, where a stiffening of the phonon mode at the M point occurs with increasing $c/a$. A larger $c/a$ ratio should improve the dynamical stability of the structure, and this crossover to positive (real) phonon energies between $c/a=1.84$ and $1.95$ correlates well with the experimental plateau of $c/a$ at 1.93. The $\Gamma$ point instability is insensitive to changes in $c/a$ (see Supplemental Material for calculated phonon dispersion curves) but an additional expansion (+25\%) of the unit cell volume can eliminate all of the negative phonon energies. However, we note that atomic volumes are fixed in the range $21-22$ {\AA}$^3$ for the bulk allotropes ($\alpha, \beta, \gamma$) and a large volume increase was not found experimentally for the hcp-like layers.

The combination of (i) the power law ($B\approx-1$) behaviour of the lattice parameters across the entire $t_{\mathrm{U}}$ range and (ii) the strong right-handed asymmetry on every specular (00.2) peak implies that there is a continuous $c$-axis expansion normal to the sample surface. If the critical thickness for relaxation is vanishingly small it is likely that, in the thickest samples, a continuum of lattice parameters exists with unit cells ranging from near-isotropic (e.g. $c/a=1.63$) at the interface to highly anisotropic ($c/a=1.93$) at the surface. 

Extrapolating the Ir/U power laws of Fig.~\ref{fig:IrCu_CValues} back to 70 {\AA} (to represent the thinnest films in published literature) gives $a=3.08$ {\AA} and $c=5.35$ {\AA} ($c/a=1.74$), which is consistent with a match of $a_\textrm{Ir/U}=3.135$ {\AA} at the interface (Section III). The significant difference between $a_\textrm{Cu/U} = 3.03(1)$ {\AA} and the expected value of $a=2.95$ {\AA} (from Fig.~2) in the thinnest Cu films may be driven by interfacial disorder (the Cu roughness and uranium rocking curve FWHM are double those of Ir), an alternative epitaxial match (such as a 30$^\circ$ relative rotation between the uranium and Cu) or other microstructural differences in the buffer (different crystallite shapes). We note that evaporation of uranium onto single crystal Cu(111) in the sub-monolayer limit has shown that low energy uranium atoms can embed themselves in the Cu surface by ejecting several substrate atoms and this may cause significant intermixing at the interface \cite{Feng2021ComparativeSurfaces}. The surface stability of Ir(111) upon uranium deposition has not yet been studied in a similar way.


Table~\ref{tableparameters} combines the results of this work with the lattice parameters of hcp-U found in literature, highlighting the general agreement between all theoretical calculations and the significant differences in experimentally derived parameters. It is interesting to note that the pseudo-bulk parameters of $c=5.64\pm0.01$ {\AA} (for U on the Ir buffer) and $c=5.65\pm0.02$ {\AA} (for U on the Cu buffer) extracted from the asymptotes of the power laws are comparable to the $c$-axis of $\beta$-U (5.656 {\AA}) \cite{Lawson1988}. Although it is not possible to determine from the electron diffraction profiles in References \cite{Chen2019DirectW110} and \cite{Molodtsov1998DispersionMetal} whether the hcp-U layers on W(110) are surface reconstructions or true single crystals, a unit cell with $a=3.15$ {\AA} or $c=5.4$ {\AA} for $t_U =70-80$ {\AA} does not seem unreasonable in the context of this paper.



\begin{table}[t]
\begin{ruledtabular}
\begin{tabular}{cccccc}
 Surface& $t_{\mathrm U}$ (\AA)&$a$ ({\AA})&$c$ (\AA)&$c/a$ & Ref.\\
\hline
W(110)& 70 & 3.15(1) & 5.1
& 1.62 & \cite{Chen2019DirectW110} \\
W(110)& 80 & 3.5(5) & 5.4(6)
& $1.5(3)$ & \cite{Berbil-Bautista2004ObservationSpectroscopy} \\
W(110)& 80 & 3.2(1) & N/A
& N/A & \cite{Molodtsov1998DispersionMetal}\\
Gd(00.1)& 500 & 2.96(2) & 5.625(5) & 1.90(1) & \cite{Springell2008ElementalFilms}  \\
\hline 
Theory &  & 2.97 & 5.35
& 1.80 & \cite{Berbil-Bautista2004ObservationSpectroscopy}  \\
Theory & & 3.01 &  5.48
& 1.82 & \cite{Schonecker2012FerromagneticPrediction}  \\
Theory & & 2.983 & 5.478 
& 1.836 & \cite{Wu2020Spin-dependentUranium}  \\
\hline
Theory & & 2.98 &  5.48
& 1.84 &  [*]\\
Cu\{111\} & 201 & 3.03(1) & 5.47(1)
& 1.804(7) &  \\
Ir\{111\} & 175 & 2.98(2) & 5.507(3)
& 1.85(2) &   \\
Cu\{111\} & 2800 & N/A & 5.635(1)
& N/A &  \\
Ir\{111\} & 2800 & 2.919(1) & 5.626(2)
& 1.928(1) &   \\
\end{tabular}
\end{ruledtabular}
\caption{Summary of experimental and theoretical (relaxed) lattice parameters of hcp-U from published literature and from this work (bottom three entries). Error bars are given in brackets when available and refer to least significant digit. [*] is the result from Section II of this paper. \label{tableparameters}}

\end{table}

\section{Transitions into $\alpha$-U}

\begin{figure*}[t]
\includegraphics[width=0.98\textwidth]{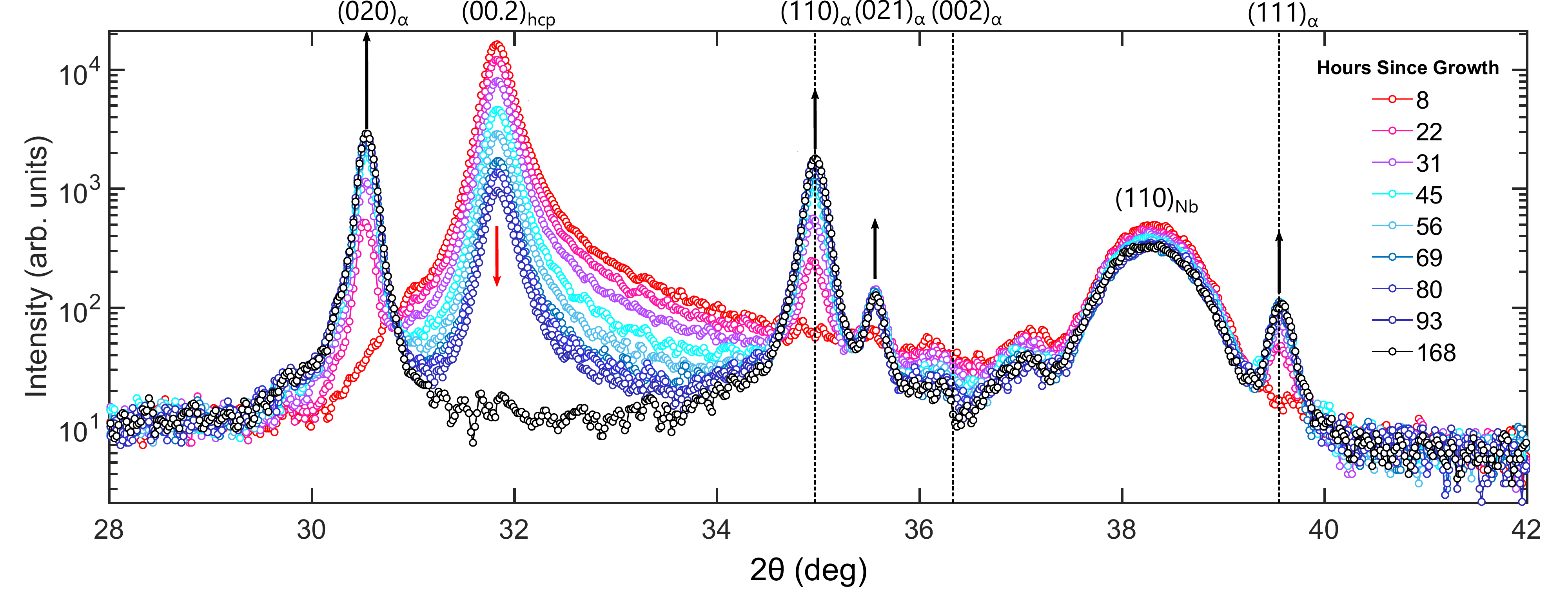}
\caption{\label{fig:overalldecay}Specular scans of a 1400 {\AA} hexagonal close-packed uranium film grown onto a polycrystalline Cu\{111\} surface over the first 7 days following growth. Arrows indicate the direction of growth/decay of the Bragg peaks. The dashed line represents the position of bulk (002)$_\alpha$. These data have not been corrected for absorption of the X-ray beam. }
\end{figure*}

The samples from Section IV transitioned into $\alpha$-U on relatively slow, thickness dependent timescales (from several hours up to several weeks). This gradual conversion was probed via X-ray diffraction. In the following discussion we use subscripts $(hkl)_\alpha$ and $(hk.l)_{hcp}$ to distinguish the different crystallographic planes and directions for the two phases. Fig$.$ \ref{fig:overalldecay} shows the time-dependent XRD data for the $t_{\mathrm{U}} = 1400$ {\AA} sample grown on Cu, where the $(00.l)_{hcp}$ peak intensities decrease monotonically with time and the $\alpha$-U peaks concomitantly increase in intensity over a period of 7 days. The functional forms of the changes in integrated intensity of the $(00.2)_{hcp}$, $(020)_{\alpha}$ and $(110)_{\alpha}$ specular peaks are well-fitted by exponential functions with similar time constants. If the transitions were propagating layer-by-layer (e.g. if the transition nucleated at the buffer interface and travelled up toward the surface) the {\it hcp} lineshape would be altered and/or shifted during the transition. As neither the $(00.2)_{hcp}$ peak center nor the asymmetry parameters change during the transition, it is likely that transitions are occurring within individual, narrow, column-like grains.  

Similar diffraction data were obtained for all other uranium thicknesses, revealing that the final $\alpha$-U reflections present in the specular direction change as a function of $t_{\mathrm{U}}$ and that no diffraction peaks arise in the specular direction other than those related to $\alpha$-U. No changes in the $\alpha$-U peak intensities were observed once the hcp-U intensity had been fully depleted, suggesting that grains of $\alpha$-U are in their final state and do not reorient post-transition. To graphically represent the complex combinations and proportions of the final $\alpha$-U orientations, the texture coefficient, $P_{hkl}$ has been used as an approximate measure of the relative volume of grains belonging to each orientation. $P_{hkl}$ is defined here as
\begin{equation}
    P_{hkl}=I_{hkl}/\Sigma_{hkl}I_{hkl}
\end{equation}
where $I_{hkl}$ is the integrated intensity of the peak in a coupled $2\theta$-$\omega$ scan when aligned to the maximum of the respective rocking curve, and the summation is over all diffraction peaks in the range $2\theta=25-50^\circ$. This range encompasses the first allowed reflections for the $\alpha$-U \{001\}$_{\alpha}$, \{110\}$_{\alpha}$, \{021\}$_{\alpha}$, \{001\}$_{\alpha}$ and \{111\}$_{\alpha}$ families and excludes the \{100\}$_{\alpha}$ orientation ($2\theta_{200}=65.34^\circ$) which was not found in any sample.

\begin{figure}[t]
\includegraphics[width=0.4\textwidth]{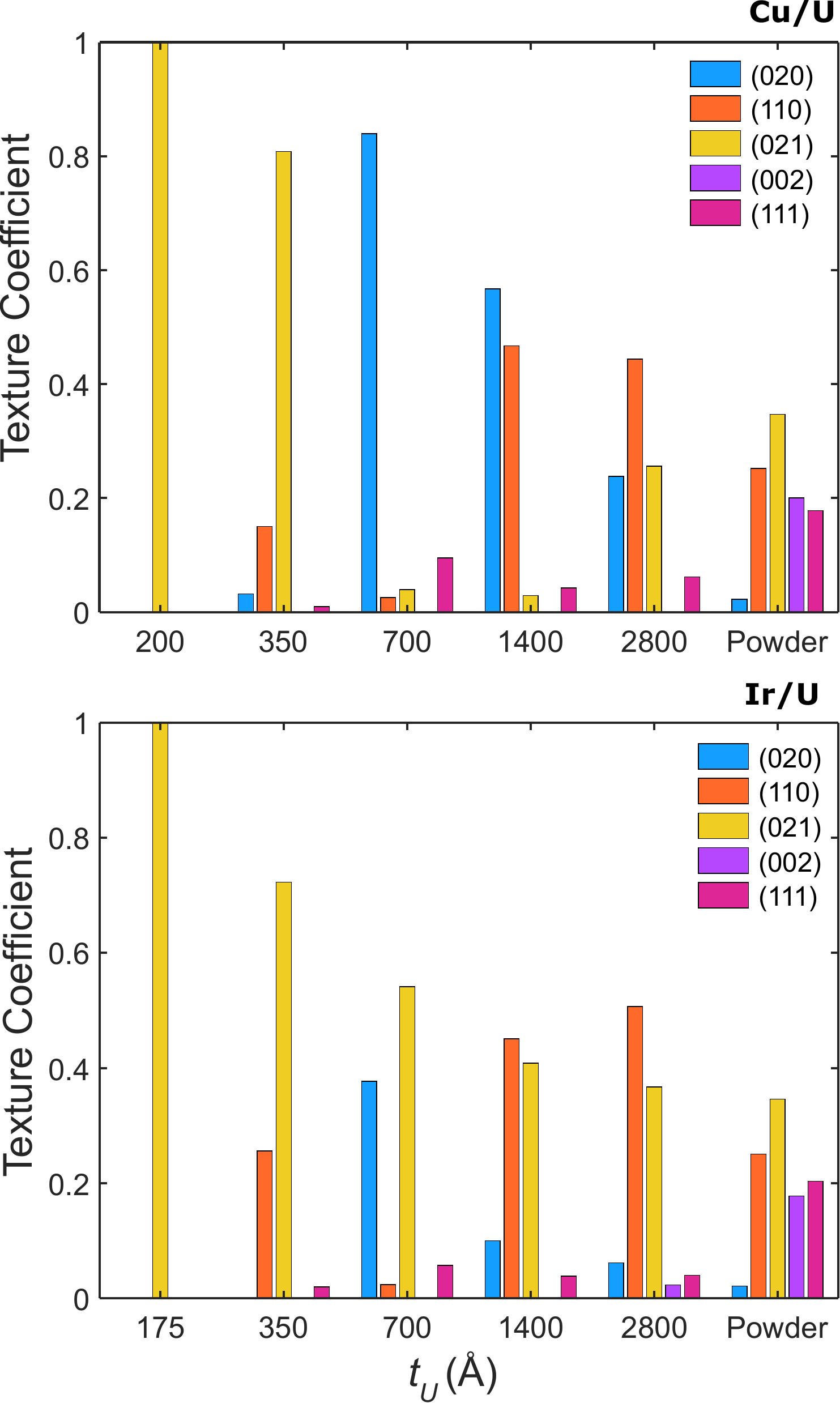}
\caption{\label{fig:texture}Grain orientations in the uranium layers found in each bilayer system post-transition, represented as an experimental texture coefficient to be compared to a theoretical powder pattern. Each system has transitioned from an initial $(00.1)_{hcp}$ oriented structure into various orientations of orthorhombic $\alpha$-U shown in the legend.}
\end{figure}

The texture coefficients of the $\alpha$-U layers in both the Ir and Cu systems are displayed in Fig.~\ref{fig:texture} as a function of $t_{\mathrm{U}}$. The $(111)_{\alpha}$ reflection is consistently suppressed compared to powder and the $(002)_{\alpha}$ reflection is absent in all but the thickest Ir/U bilayers. The only ground state available to the thinnest films in both series is the $(021)_{\alpha}$ orientation, while the mid-range layers can access the $(110)_{\alpha}$ and $(010)_{\alpha}$ orientations. Notably, the 700 {\AA} Cu/U film becomes almost pure (010)$_\alpha$. The thickest films form more textured $\alpha$-U with no single dominant orientation. It is important to note that the texture coefficient is a semi-quantitative representation, as possible extinction and absorption effects have not been taken into account. For Cu-K$_\alpha$ radiation incident at 15$^\circ$ degrees, the $1/e$ attenuation length is $\sim4000$ {\AA} and the absorption effects should be small for the tested film thicknesses. We also note that an orientation dependent mosaic texture is present and, if the rocking curve intensity is combined multiplicatively with the specular intensity, the texture coefficients of the (021)$_\alpha$ and (110)$_\alpha$ orientations in the mid-range samples are altered (maximally by $\pm0.2$). These caveats do not impact the overall results of Fig.~\ref{fig:texture}.

\begin{table}[]
\centering
\begin{ruledtabular}
\begin{tabular}{ccc}
Match Direction   &  Comparable Spacing & Distance ({\AA}) \\
\hline

(00.1)$_{hcp}$   & $d_{00.1}$ & 5.47(1)$^*$-5.635(1) \\
(111)$_\alpha$   & 2$d_{111}$ & 4.558 \\
(001)$_\alpha$   &   $d_{001}$ & 4.955 \\
(021)$_\alpha$   &  2$d_{021}$ & 5.049  \\
(110)$_\alpha$    &     2$d_{110}$ & 5.132    \\
(010)$_\alpha$     &    $d_{010}$ & 5.869  \\
\end{tabular}
\end{ruledtabular}
\caption{Comparison of the $c$-axis length in the hcp-U bilayers to similar interplanar spacings in various orientations of $\alpha$-U. (*) denotes the parameters from the 201 {\AA} Cu/U sample, but thinner layers may have smaller $c$ axes. }
\label{tab:latticespacing}
\end{table}


\begin{figure}[t]
\includegraphics[width=0.48\textwidth]{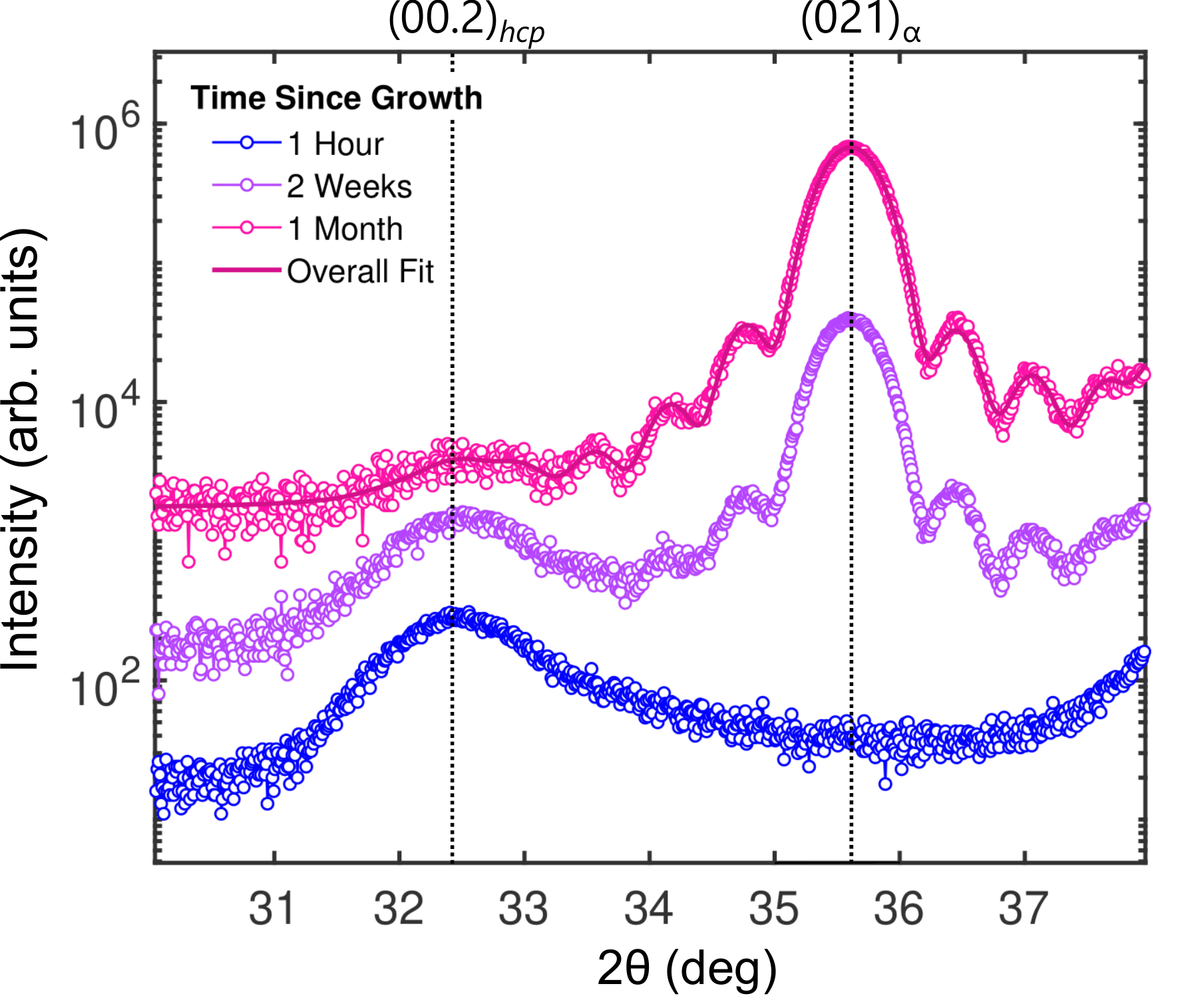}
\caption{\label{fig:Ir_175_DECAY} Specular scans of the Ir/U system with an XRR-derived thickness of $t_{\mathrm{U}} = 175$ {\AA} in the month after growth. An $(021)_\alpha$ reflection ($2d_{021}=5.037$ {\AA}) with Laue fringes emerges at the expense of the $(00.2)_{hcp}$ peak ($c=5.507$ {\AA}). The cut-off intensity on the right hand side of the figure is a broad (110) reflection from the Nb cap.}
\end{figure}

Further insight into the microstructure of these samples can be gained from the time-resolved diffraction scans of thinner films. Fig.~\ref{fig:Ir_175_DECAY} shows several specular scans of the $t_{\mathrm{U}} = 175$ {\AA} Ir/U bilayer taken over the month following growth. The emergence of Laue fringes indicates that the original $(00.1)_{hcp}$ layer has become a coherent (021)$_\alpha$ layer with a sharp Ir-U interface. These fringes were presumably suppressed in the initial hcp films by the continuously changing lattice parameters, as discussed in Section IV. As the periodicity of the fringes is related to the number of scattering planes perpendicular to the surface normal, the thickness of the crystalline ordered volume can be extracted from a fit to the oscillations. For the Ir/U system with $t_{\mathrm{U}} = 175$ {\AA}, we find $t_{\mathrm{Laue}}=155\pm 2$ {\AA}. For the Cu/U system with $t_{\mathrm{U}} = 201$ {\AA} we find $t_{\mathrm{Laue}}=189\pm6$ {\AA}. While both of these values are slightly smaller than the XRR derived thicknesses, the close agreement supports the previous picture of columns extending through the full thickness of the films. Laue fringes in the $t_U=175-700$ {\AA} high angle scans suggest that coherent columns are maintained up to layer thicknesses of at least 700 {\AA} for both the Cu and Ir buffered systems. 

We finally return to the observation that the hcp-U intensity in the thickest films rapidly diminishes, while the intensity in the thinnest films persists for several weeks. This suggests that there are kinetic barriers to these transitions that are dependent on some microscopic parameter(s) that changes with thickness, e.g. the $c/a$ ratio and the unit cell volume. The competition between the $(010)_{\alpha}$, $(110)_{\alpha}$ and $(021)_{\alpha}$ ground states and absence of the $(002)_{\alpha}$ reflection can be naively explained by Table II, which shows that smaller $c_{hcp}$-axes (i$.$e$.$ thinner hcp layers) are more closely aligned with the $(021)_{\alpha}$ and $(110)_{\alpha}$ interplanar spacings and larger $c_{hcp}$-axes (i$.$e$.$ thicker layers) lie closer to the $b$-axis of $\alpha$-U. The in-plane match directions are more complex and will be discussed in Section VI, along with potential lattice distortions which drive the transitions into the three primary $\alpha$-U orientations of Fig.~\ref{fig:texture}.




\section{Transition Pathways}
\begin{figure}[t]
\includegraphics[width=0.48\textwidth]{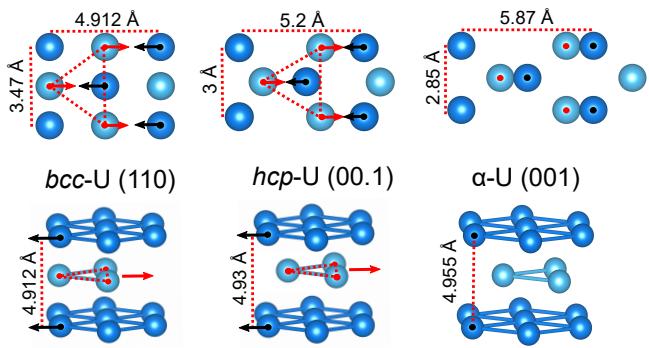}
\caption{\label{fig:Axe}Illustration of the Axe mode from Ref.~\cite{Axe1994StructureMetal}, showing transformations induced by [110] transverse acoustic displacements in a $bcc$-U lattice. Red and black arrows represent in-plane displacements of hexagonal `sheets' in opposite directions. In the upper images, the light blue layer of atoms exist in the plane below the dark blue atoms. These alternate planes (i.e. the (110) planes in the $bcc$ structure) are shifted in opposite directions along the [1-10] axis. The in-plane aspect ratio is stretched from $\sqrt{2}$ to $\sqrt{3}$, giving rise to an intermediate hcp (00.1) structure (P6$_3$/$mmc$). Stretching the ratio further and continuing the alternate layer shifts gives the $\alpha$-U (001) structure. The estimated hcp-U lattice parameters are taken as the midpoint between the two endpoint structures.}
\end{figure}

The first attempt to geometrically relate the $\alpha$-U and hcp-U structures can be found in Ref.~\cite{Axe1994StructureMetal}, where an hcp structure arises as a by-product of a displacive transition from the body-centred cubic $\gamma$-U phase to the $\alpha$-U phase (ignoring the $\beta$-U structure). The Axe mechanism, illustrated in Fig.~\ref{fig:Axe}, maps a single hcp-U unit cell onto a single $\alpha$-U unit cell via the freeze-in of a transverse acoustic phonon. The displacement of alternate $\gamma$-U (110) layers along the [1-10]$_\gamma$ direction generates an intermediate, isotropic hexagonal close-packed structure ($c/a=1.64$) with its $c$-axis parallel to the \textit{c}-axis of $\alpha$-U. 

The absence of the $(001)_{\alpha}$ orientation in the transition products (Section V) and the large $c/a$ ratios (Section IV, $c/a=1.80$-$1.93$) suggest that the Axe mode is not the mechanism governing the transitions in these metastable Ir/U and Cu/U systems. For example, to convert the $t_U=2800$ {\AA} $(00.1)_{hcp}$ unit cell into $(001)_{\alpha}$, a $c$-axis contraction of 12\% would be required instead of the 0.9\% expansion shown in Fig.~\ref{fig:Axe}. There are no other displacive mechanisms which can convert a single, anisotropic ($c/a=1.93$) unit cell of hcp-U into $\alpha$-U with strains of under 10\%. Any instabilities pointing toward the Axe distortion have not yet been observed in either bulk $\gamma$-U or $\alpha$-U. It is clear from Section V that several other pathways linking the hcp structure to the orthorhombic ground state do exist, and that there are well-defined relationships between the initial hcp lattice parameters and the final orientations of $\alpha$-U.  

\begin{figure}[t]
\includegraphics[width=0.48\textwidth]{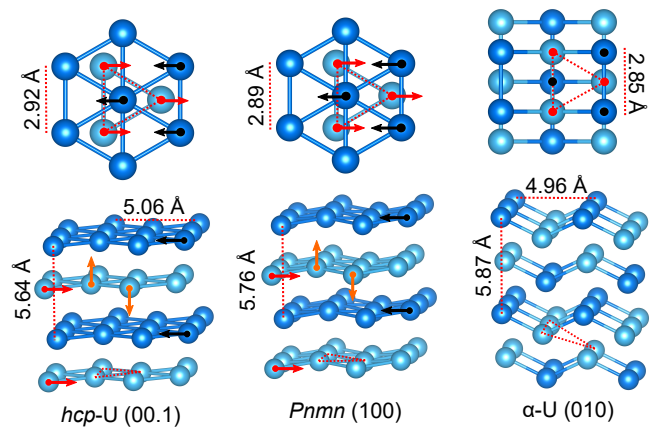}
\caption{\label{fig:020transform}Schematic of the transformation from hcp-U (00.1) ($a=2.92$ {\AA} and $c=5.63$ {\AA}) to $\alpha$-U (010) via the {\it Pnmn} structure. Red and black arrows represent in-plane displacements of the `sheets' (in opposite directions) and orange arrows represent the out-of-plane displacements of alternate atoms within these sheets. In the upper left and centre images, the lighter blue layer of atoms exist in the plane below the darker blue atoms.}
\end{figure}

Fig.~\ref{fig:020transform} shows an alternative, two-step phase transformation from (00.1)$_{hcp}$ into (010)$_\alpha$. The pathway was determined using COMSUBS, a sub-program from the ISOTROPY suite which finds pathways between parent and daughter phases via a shared symmetry subgroup \cite{Stokes2002ProcedureSolids,Stokes2021ISODISTORTSuite}. Searches were initially limited to combinations of strain modes with $\pm$10\% of the parent lattice parameters, displacive modes with a maximum atomic displacement of 1 {\AA} and supercells comprising up to six primitive cells of the parent structure.  Fig.~\ref{fig:020transform} shows the only hcp-to-$\alpha$ pathway available within these limits for a parent unit cell with $a=2.92$ {\AA} and $c=5.63$ {\AA}. 

In the proposed (00.1)$_{hcp}$ to (010)$_\alpha$ transition, a supercell comprised of two primitive hcp cells is mapped onto an intermediate {\it hcp}-like orthorhombic (Pnmn) structure via the motion of alternate `sheets' in opposite directions (red and black arrows) and the shearing of alternate rows of atoms within these sheets in directions parallel to $c_{hcp}$ (orange arrows). Further shearing, bonds breaking and reforming gives rise to the corrugated layers characteristic of $\alpha$-U. Strain modes act on the supercell throughout, with the overall transition resulting in a 4.2~\% expansion of the original $c_{hcp}$ axis, a 2.3~\% contraction of the $a_{hcp}$ axis and a 3.4~\% contraction along the [11.0]$_{hcp}$ axis. These strains are significantly lower than the 12~\% distortion required by the Axe mode. If an entire sample volume passed through this transition at a slow and linear rate, the initial specular peak near $2\theta=32^\circ$ would shift gradually toward $2\theta=30.44^\circ$ for bulk (020)$_\alpha$, while dropping rapidly in intensity. No other peaks would appear in the specular direction during the transition, consistent with experiment. If the Axe model proceeded in the same steady fashion, the spectra would be dominated instead by a shift of the specular peak toward $2\theta=36.23^\circ$ for bulk (001)$_\alpha$, coupled with a negligible change in peak intensity. 

It is interesting to note that the atomic motions indicated in Fig.~\ref{fig:020transform} also correspond to the displacements associated with the eigenvectors of the imaginary modes in Fig.~\ref{fig:PhononDispersion}. Regardless of $q$-vector, the phonon modes are related to optical displacements of the two atoms in the hcp cell in opposite directions in the $x$ and $y$ directions, relative to the basal plane. It is possible that temporary stability of an hcp-like structure is attained via small distortions along these axes, whilst also paving the way for the gradual transitions back into $\alpha$-U. 

\begin{figure}[t]
\includegraphics[width=0.48\textwidth]{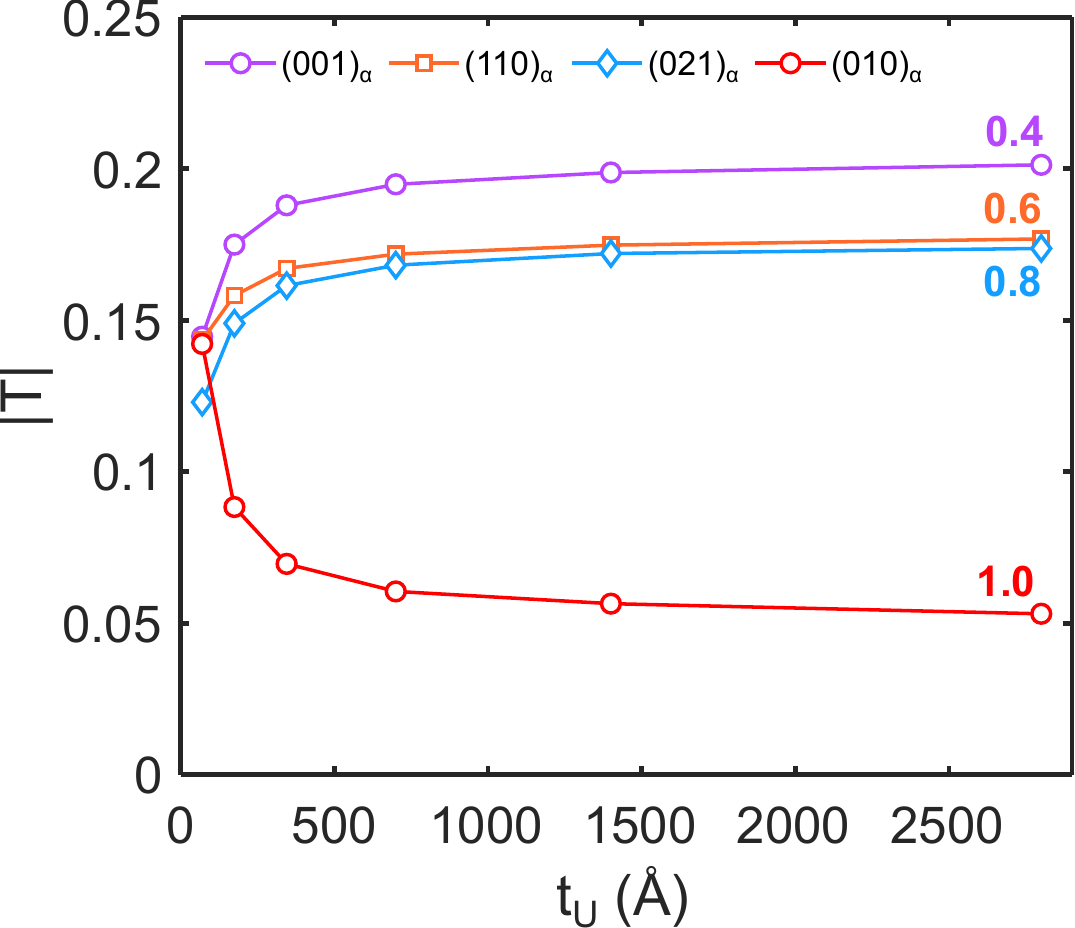}
\caption{\label{fig:StrainTensor}Thickness dependence of the magnitude of the percentage strain tensor, $T$ for various transition pathways. Attached to each line is the maximum atomic displacement within the transition in {\AA}. Data points correspond to values of $a$ and $c$ extracted from the power law fit to the experimentally derived Ir/U bilayer lattice parameters.}
\end{figure}

In order to explore alternative pathways, the COMSUBS boundary conditions were relaxed to allow atomic displacements up to 1.4 {\AA} and strains of $\pm$~15\% for various starting structures (see Supplemental Material for a summary of all pathways that do not require triclinic supercells). Each pathway has an associated strain tensor, $T$ of the form [A\% B\% C\%] which describes the expansion or contraction of the supercell along the principal strain directions. The magnitude of this strain tensor, $\mid T\mid=\sqrt{A^2+B^2+C^2}$ has been plotted in Fig.~\ref{fig:StrainTensor} as a function of thickness for each of the (001)$_\alpha$, (110)$_\alpha$, (021)$_\alpha$ and (010)$_\alpha$ transitions. The maximum atomic shuffle in each transition is attached to the respective line. As the transitions into (021)$_\alpha$ (requiring a 4 unit cell supercell) and (110)$_\alpha$ (requiring a 2 unit cell supercell) are too complex to show in the main text, please see Supplemental Material for 2D illustrations and animated figures.

On the basis of the reduced atomic shuffle and strain, the (021)$_\alpha$ transition should be favoured in the thinnest films with the lowest $c/a$ ratios. For unit cells with larger $c/a$ ratios, there is a rapid drop-off in the strain required for the (010)$_\alpha$ transition but the larger atomic shuffle may promote competition with the (021)$_\alpha$ and (110)$_\alpha$ transitions. These results are consistent with experimental observations, but a more detailed study using density functional theory would allow the quantitative calculation of the height of the energy barrier to each transition. The reaction kinetics could also be explored by storing the samples at cryogenic temperatures and tracking any increase in the hcp lifetime. 

Finally we draw links to Ref.~\cite{Ward2008TheUranium}, where uranium metal deposited onto Gd(00.1) at 600$^\circ$C formed domains of (021)$_\alpha$ and (110)$_\alpha$ with definite crystalline order, but no discernible match to the hexagonal buffer layer. If the system had initially grown as a two-domain hcp layer with a 30$^\circ$ relative rotation between domains, as in Ref.~\cite{Springell2008ElementalFilms}, a conversion of Domain 1 (U[10.0]//Gd[10.0]) into (110)$_\alpha$ and Domain 2 (U[21.0]//Gd[10.0]) into (021)$_\alpha$ via the mechanisms suggested here gives the orientations of $\alpha$-U found in Ref.~\cite{Ward2008TheUranium}. It is likely that a metastable, well-ordered hcp layer transitioned into $\alpha$-U via the same pathways seen in these Cu/Ir buffered systems, culminating in the unexplained epitaxial relationships.


\section{Conclusions}
We have shown that metastable hcp-like layers of uranium metal can be grown onto \{111\}-textured Cu and Ir surfaces at room temperature, giving a highly anisotropic `relaxed' structure with experimental lattice parameters ($a=2.92$ and $c=5.63$ {\AA}, $c/a=1.93$) that deviate significantly from the relaxed theoretical hcp values ($a=2.98$ and $c=5.48$ {\AA}, $c/a=1.84$). These films gradually and predominantly relax into the (010), (110) and (021) orientations of $\alpha$-U. We have suggested several two-step mechanisms for these transitions and encourage more quantitative theoretical studies of the pathways. The relatively long lifetimes of the thinner hcp-like layers open the door to further studies of the electronic and magnetic properties of this metastable structure, which could reveal magnetic or superconducting order at lower temperatures. These results emphasize both the fascinating complexity of this heavy elemental metal, and the power of thin film growth techniques to explore novel crystal symmetries. 

\begin{acknowledgments}
This research was supported by the Engineering and Phys. Sciences Research Council (EPSRC), UK, through the Centre for Doctoral Training in Condensed Matter Physics (CDT-CMP) grant no. \textsc{EP/L015544/1}, and the National Nuclear User Facility for Radioactive Materials Surfaces (NNUF-FaRMS), grant no. \textsc{EP/V035495/1}. We acknowledge highly useful discussions with D. Chaney.

\end{acknowledgments}

\end{document}